\documentclass{jfm}
\usepackage[T1]{fontenc}
\usepackage[latin9]{inputenc}
\usepackage{soul}
\usepackage{xcolor}
\usepackage{graphicx}
\usepackage{amssymb}
\usepackage[authoryear]{natbib}

\makeatletter

\providecolor{lyxadded}{rgb}{0,0,1}
\providecolor{lyxdeleted}{rgb}{1,0,0}

\ifCUPmtlplainloaded \else
  \IfFileExists{amsbsy.sty}
    {\typeout{^^JFound the 'amsbsy' package on the system, using it.^^J}
}

\def\vec#1{\mbox{\boldmath $\mathit{#1}$}}

\def\RE{\textit{Re}}
\def\Rm{\textit{Rm}}
\def\Ha{\textit{Ha}}
\def\i{\mathrm{i}}

\begin{document}

\title[Linear stability of MHD flow in a perfectly conducting rectangular duct]
{Linear stability of magnetohydrodynamic flow in a perfectly conducting 
rectangular duct}
\author[J. Priede, S. Aleksandrova and S. Molokov]
{J\ls \=A\ls N\ls I\ls S\ns P\ls R\ls I\ls E\ls D\ls E,\ns 
S\ls V\ls E\ls T\ls L\ls A\ls N\ls A\ns A\ls L\ls E\ls K\ls S\ls A\ls N\ls D\ls R\ls O\ls V\ls A\\
and\ns 
S\ls E\ls R\ls G\ls E\ls I\ns M\ls O\ls L\ls O\ls K\ls O\ls V}

\maketitle

\begin{abstract}
We analyse numerically the linear stability of a liquid metal flow
in a rectangular duct with perfectly electrically conducting walls
subject to a uniform transverse magnetic field. A non-standard three
dimensional vector stream function/vorticity formulation is used with
Chebyshev collocation method to solve the eigenvalue problem for small-amplitude
perturbations. A relatively weak magnetic field is found to render
the flow linearly unstable as two weak jets appear close to the centre
of the duct at the Hartmann number $\Ha\approx9.6.$ In a sufficiently
strong magnetic field, the instability following the jets becomes
confined in the layers of characteristic thickness $\delta\sim\Ha^{-1/2}$
located at the walls parallel to the magnetic field. In this case
the instability is determined by $\delta,$ which results in both
the critical Reynolds and wavenumbers numbers scaling as $\sim\delta^{-1}.$
Instability modes can have one of the four different symmetry combinations
along and across the magnetic field. The most unstable is a pair of
modes with an even distribution of vorticity along the magnetic field.
These two modes represent strongly non-uniform vortices aligned with
the magnetic field, which rotate either in the same or opposite senses
across the magnetic field. The former enhance while the latter weaken
one another provided that the magnetic field is not too strong or
the walls parallel to the field are not too far apart. In a strong
magnetic field, when the vortices at the opposite walls are well separated
by the core flow, the critical Reynolds and wavenumbers for both of
these instability modes are the same: $\RE_{c}\approx642\Ha^{1/2}+8.9\times10^{3}\Ha^{-1/2}$
and $k_{c}\approx0.477\Ha^{1/2}.$ The other pair of modes, which
differs from the previous one by an odd distribution of vorticity
along the magnetic field, is more stable with approximately four times
higher critical Reynolds number. 
\end{abstract}

\section{Introduction}

Understanding instabilities in magnetohydrodynamic (MHD) flows in
ducts is of great importance for liquid metal flows in blankets for
fusion reactors \citep{Buh07}. Blankets consist of rectangular ducts
in which the liquid metal flows in a high, transverse magnetic field
of between 5 and 10~T. The aim of these devices is to cool plasma
chamber and to breed and to remove tritium. This can be assisted by
mixing of the flow by turbulence if it can be sustained in the presence
of a magnetic field. A high magnetic field is known to damp turbulence
by means of the Lorentz force. At the same time, the magnetic field
can also affect the base velocity profile in such a way as to create
inflection lines \citep{Kaku64} and even jets \citep{Hunt65} thus
making the flow more unstable. These two competing effects balancing
each other on a certain length scale result in relatively simple asymptotics
for the instability threshold. The most dangerous perturbations are
usually associated with the largest length scale on which the magnetic
damping becomes comparable with the viscous one. This happens in the
so-called parallel layers with the relative thickness $\sim\Ha^{-1/2}$
\citep{Bra60}. Linear stability of these layers for a duct with insulating
walls \citep{She53} has been considered in a quasi-two-dimensional
approximation by \citet{Pot07}, who found the critical Reynolds number
$\RE_{c}\approx4.8\times10^{4}\Ha^{1/2}.$ It is an extremely high,
however, typical value for the linear stability of exponential velocity
profiles, which have the critical Reynolds number around fifty thousand
based on the boundary layer thickness \citep{Drazin-Reid-81}. This
high threshold is of little practical relevance because the instability
for exponential velocity profile is known to be subcritical \citep{Hocking-75}.
This is the case also for the stability of Hartmann layer \citep{Lock55},
which is subcritical too \citep{LifSht79,MorAlb03} with the experimentally
found Reynolds number for the onset of turbulence in straight and
annular ducts of rectangular cross-sections being respectively around
$225\Ha$ \citep{Mur53,BroLyk67} and $380\Ha$ \citep{MorAlb04}.
Marginal turbulent flow states have been observed by \citet{ShaGer10}
significantly below the linear stability threshold in numerical simulations
of insulating duct flow subject to a transverse magnetic field of
moderate strength.

The stability of MHD flows strongly varies with the electrical conductivity
of the duct walls. For example, Hunt's flow, which develops in a rectangular
duct when the walls perpendicular to the magnetic field are perfectly
conducting while the parallel ones are insulating, has a relatively
low linear stability threshold $\RE_{c}\approx91\Ha^{1/2}$ and $\bar{\RE_{c}}\approx112$
based on the maximum and average velocities, respectively \citep{PAM10}.
The low stability of Hunt's flow is due to two strong jets, which
develop in a sufficiently strong magnetic field along the insulating
walls and attain a velocity $\sim\Ha$ relative to that of the core
flow \citep{Hunt65}. Although the relative velocity of jets reduces
as $\sim\Ha^{1/2}/c$ with the increase of the wall conductance ratio
$c\gtrsim\Ha^{-1/2}$ \citep{Wal81}, weak jets with the relative
velocity $O(1)$ still persist at the parallel walls also in the limit
of perfectly conducting duct \citep{Ufl61,ChaLun61}. The presence
of jets with inherent inflection points suggests that this flow may
also be highly unstable similar to Hunt's flow. It is the aim of the
present study to investigate linear stability of this flow, which
is the last basic MHD duct flow configuration whose linear stability
may be not only of theoretical but also of experimental relevance.
We first investigate the case of square duct and find the high-field
asymptotics of the instability threshold which are then generalized
to arbitrary aspect ratios. 

The paper is organised as follows. The problem is formulated in $\S$\ref{sec:prob}
below. In $\S$\ref{sec:res} we present and discuss numerical results
for a square duct in a vertical magnetic field as well as for ducts
with various aspect ratios in both vertical and horizontal magnetic
fields. The paper is concluded with a summary and comparison with
experimental results in $\S$\ref{sec:sum}. Our non-standard vector
stream function/vorticity formulation is described in Appendix \ref{sec:app}.

\begin{figure}
\begin{centering}
\includegraphics[bb=60bp 60bp 400bp 280bp,clip,width=0.6\columnwidth]{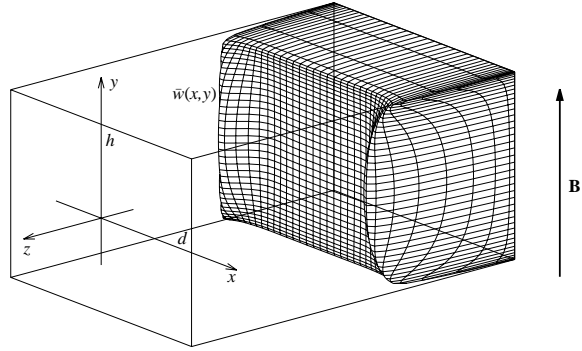} 
\par\end{centering}

\caption{\label{fig:sketch}The base flow profile in a rectangular duct with
perfectly conducting walls subject to a strong vertical magnetic field
for $\Ha=100.$ }
\end{figure}

\section{\label{sec:prob}Formulation of the problem}

Consider a flow of an incompressible viscous electrically conducting
liquid with density $\rho,$ kinematic viscosity $\nu$ and electrical
conductivity $\sigma$ driven by a constant gradient of pressure $p$
applied along a straight duct of rectangular cross-section with half-width
$d$ and half-height $h$ subject to a transverse homogeneous magnetic
field $\vec{B}.$ The walls of the duct are assumed to be perfectly
electrically conducting and the field may be applied across either
the width or the height of the duct.

The velocity distribution of the flow is governed by the Navier-Stokes
equation\begin{equation}
\partial_{t}\vec{v}+(\vec{v}\cdot\vec{\nabla})\vec{v}=-\rho^{-1}\vec{\nabla}p+\nu\vec{\nabla}^{2}\vec{v}+\rho^{-1}\vec{f},\label{eq:NS}\end{equation}
 where $\vec{f}=\vec{j}\times\vec{B}$ is the electromagnetic body
force involving the induced electric current $\vec{j},$ which is
governed by the Ohm's law for a moving medium \begin{equation}
\vec{j}=\sigma(\vec{E}+\vec{v}\times\vec{B}).\label{eq:Ohm}\end{equation}
 The flow is assumed to be sufficiently slow so that the induced magnetic
field is negligible relative to the imposed one, which supposes the
magnetic Reynolds number $\Rm=\mu_{0}\sigma v_{0}d\ll1,$ where $\mu_{0}$
is the permeability of vacuum and $v_{0}$ is the characteristic velocity
of the flow. In addition, we assume that the characteristic time of
velocity variation is much longer than the magnetic diffusion time
$\tau_{m}=\mu_{0}\sigma d^{2},$ which allows us to use the quasi-stationary
approximation leading to $\vec{E}=-\vec{\nabla}\phi,$ where $\phi$
is the electrostatic potential \citep{Rob67}. The velocity and current
satisfy the mass and charge conservation $\vec{\nabla}\cdot\vec{v}=\vec{\nabla}\cdot\vec{j}=0.$
Applying the latter to the Ohm's law (\ref{eq:Ohm}) yields \begin{equation}
\vec{\nabla}^{2}\phi=\vec{B}\cdot\vec{\omega},\label{eq:phi}\end{equation}
 where $\vec{\omega}=\vec{\nabla}\times\vec{v}$ is vorticity. At
the duct walls $S$, the normal $(n)$ and tangential $(\tau)$ velocity
components satisfy the impermeability and no-slip boundary conditions
$\left.v_{n}\right\vert _{s}=0$ and $\left.v_{\tau}\right\vert _{s}=0.$
As the walls are perfectly conducting, the tangential electric current
vanishes and Ohm's law (\ref{eq:Ohm}) yields $\left.\phi\right\vert _{s}=\textrm{const}.$ 

We employ the Cartesian coordinates with the origin set at the centre
of the duct, $x$, $y$ and $z$ axes directed along its width, height
and length, respectively, as shown in figure \ref{fig:sketch}, and
the velocity defined as $\vec{v}=(u,v,w).$ The problem admits a purely
rectilinear base flow with a single velocity component along the duct
$\bar{\vec{v}}=(0,0,\bar{w}(x,y))$ which is shown in figure \ref{fig:sketch}(a)
for a strong vertical magnetic field.

In the following, all variables are non-dimensionalised by using the
maximum velocity $\bar{w}_{0}$ and the half-width of the duct $d$
as the velocity and length scales, while the time, pressure, magnetic
field and electrostatic potential are scaled by $d^{2}/\nu,$ $\rho\bar{w}_{0}^{2},$
$B=\left\vert \vec{B}\right\vert $ and $\bar{w}_{0}dB,$ respectively.
The dimensionless paramerters defining the problem are the Reynolds
number $\RE=\bar{w}_{0}d/\nu,$ the Hartmann number $\Ha=dB\sqrt{\sigma/(\rho\nu)}$
and the aspect ratio $A=h/d.$ Note that we use the maximum rather
than average velocity as the characteristic scale because the stability
of this flow, as shown in the following, is determined by the former. 

Linear stability of this flow is analysed using the same method as
in our previous study \citep{PAM10}. Since the method is based on
a non-standard vector stream function formulation, it is briefly outlined
in the Appendix. 

The problem was solved by a spectral collocation method on a Chebyshev-Lobatto
grid with even number of points defined by $2N_{x}+2$ and $2N_{y}+2$
for the $x$- and $y$-directions, where $N_{x,y}=35\cdots60$ were
used for various combinations of the control parameters to achieve
accuracy of at least three significant figures. Owing to the double
reflection symmetry of the base flow with respect to $x=0$ and $y=0$
planes, small-amplitude perturbations with different parities in $x$
and $y$ decouple from each other. This results in four mutually independent
modes, which we classify as $(o,o),$ $(o,e),$ $(e,o),$ and $(e,e)$
according to whether the $x$ and $y$ symmetry of $\hat{\psi}_{x}$
is odd or even, respectively. Our classification of modes corresponds
to the symmetries I, II, III, and IV used by \citet{TatYos90} and
\citet{UhlNag06}. Thus, four independent problems of different symmetries
are obtained in one quadrant of the duct cross-section with $N_{x}\times N_{y}$
internal collocation points. The size of matrix for each eigenvalue
problem is reduced by a factor of 16 in comparison to the original
problem. Further details and the validation of the numerical method
can be found in our previous paper \citep{PAM10}.

\section{\label{sec:res}Results}

\begin{figure}
\begin{centering}
\includegraphics[bb=130bp 90bp 370bp 260bp,clip,width=0.5\textwidth]{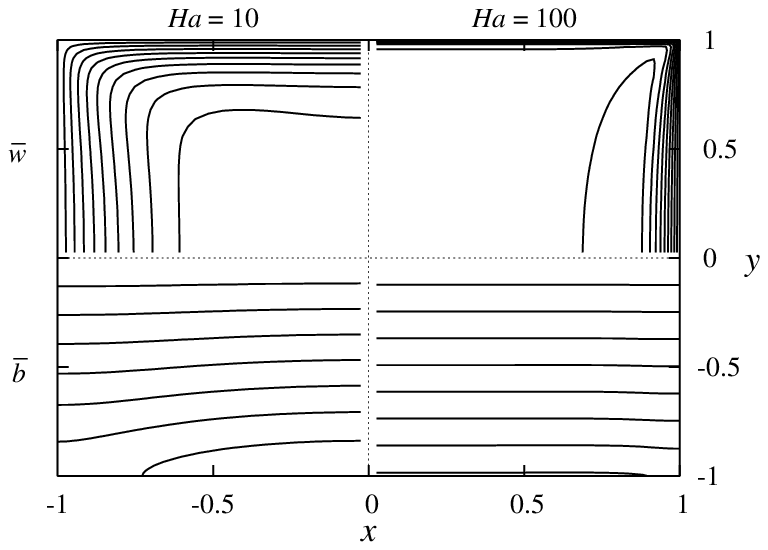}\put(-180,140){(\textit{a})}\includegraphics[width=0.5\textwidth]{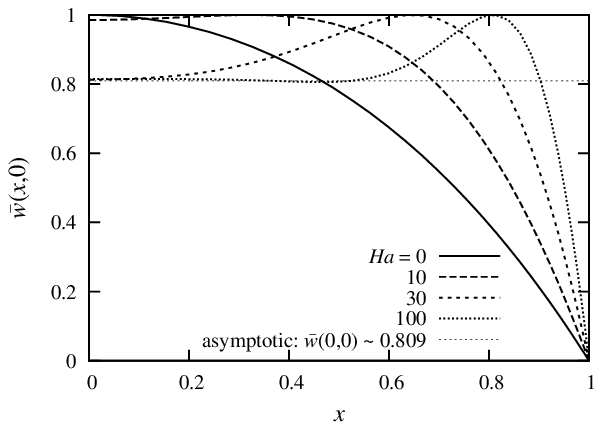}\put(-180,140){(\textit{b})} 
\par\end{centering}

\caption{\label{fig:bflow}Isolines of the base flow $(y>0)$ and electric
current lines $(y<0)$ for $\Ha=10$ $(x<0)$ and $\Ha=100$ $(x>0)$
shown in the respective quadrants of duct cross-section (a) and the
base flow velocity profiles at $y=0$ for $\Ha=0,10,30,100$ (b) in
a vertical magnetic field. }
\end{figure}

Flow in the presence of a vertical magnetic field induces a transverse
current in the bulk of the duct which, as seen in the lower part of
figure \ref{fig:bflow}(a), almost directly connects to the perfectly
conducting walls parallel to the magnetic field. However, a small
part of the current diverts in the corner regions to connect through
the Hartmann walls perpendicular to the magnetic field. This makes
the density of the transverse current and so the resulting electromagnetic
force, which opposes the constant driving pressure gradient, slightly
lower at the parallel walls than in the core region. As a result,
weak jets form along the parallel walls, where the flow becomes slightly
faster than in the core of the duct. As seen in figure \ref{fig:bflow}(a),
the formation of jets starts with a velocity minimum appearing in
the centre of the duct at $\Ha\approx10.$ With the increase of the
magnetic field, the velocity in the core becomes almost uniform, while
the jets become confined in thin layers that develop along the walls
parallel to the magnetic field and have a thickness decreasing as
$\sim\Ha^{-1/2}.$ In a strong vertical magnetic field, asymptotic
solution by \citet{Hunt65} yields the velocity maxima located in
the mid-plane of the duct at the distance \begin{equation}
\delta\approx1.915(A/\Ha)^{1/2}\label{eq:delta}\end{equation}
from the parallel wall, while the ratio of this velocity to that of
the core flow approaches a constant $0.809.$ The latter is seen in
figure \ref{fig:bflow}(b) to agree well with our numerical solution.
An interesting feature of these jets is that the extra flow rate associated
with the velocity over-shoot above the core velocity balances the
flow rate deficit at the wall, where the velocity falls below that
of the core \citep{Wil62}. Thus, the relative contribution of these
jets to the flow rate is not $\sim\Ha^{-1/2},$ as one would expect
from simple scaling arguments, but a higher-order small quantity $\sim\Ha^{-3/2},$
which is less than the flow rate correction due to the Hartmann layers
$\sim\Ha^{-1}.$ It is confirmed also by our numerical solution, which
yields the best fit of the flow rate for one quarter of a square duct
 \begin{equation}
Q\approx0.809-0.43\Ha^{-1},\label{eq:qflow}\end{equation}
where the leading-order contribution due to the core flow matches
the asymptotic solution. Note that although the correction is $\sim\Ha^{-1},$
its coefficient is not equal to that in the asymptotic solution by
\citet{Hunt65}. The difference is because the asymptotic solution
is obtained for a fixed pressure gradient, whereas numerical solution
is for a fixed maximum velocity, which has a $O(\Ha^{-1})$ higher-order
correction. Thus, the maximum velocity taken as the reference one
in this study, results in the same order correction in the core velocity
and, thus, also in the flow rate. Our choice of the maximum velocity
as the reference one is motivated by the following results, which
show that the instability in this flow is associated with the jets
at the parallel walls, which seem inherently unstable due to the inflection
points in their velocity profiles. Our results can be rescaled to
the average velocity using relation (\ref{eq:qflow}) which becomes
sufficiently accurate for $\Ha\gtrsim100.$

\begin{figure}
\begin{centering}
\includegraphics[width=0.5\textwidth]{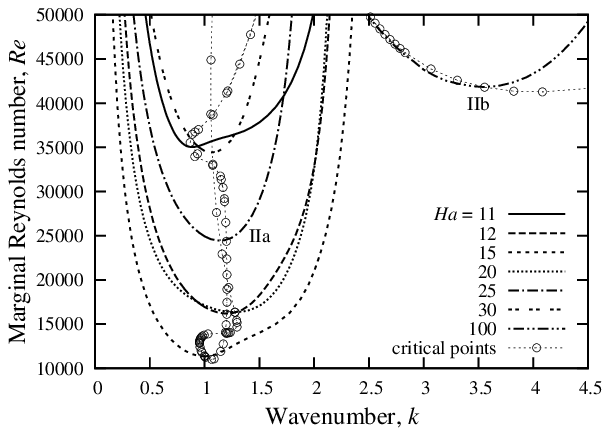}\put(-180,140){(\textit{a})}\includegraphics[width=0.5\textwidth]{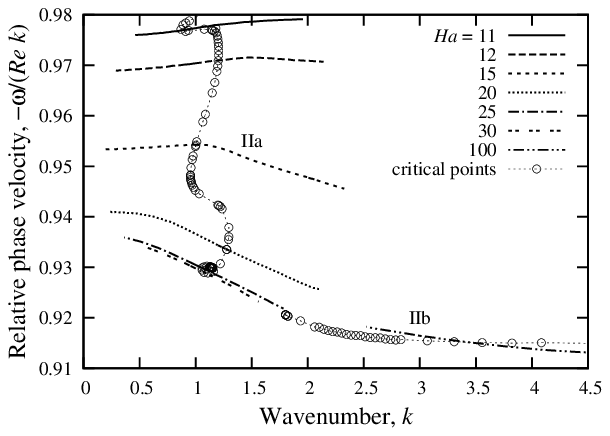}\put(-180,140){(\textit{b})} 
\par\end{centering}

\caption{\label{fig:rewk_Ge1_12}The marginal Reynolds number (a) and the relative
phase velocity (b) versus the wavenumber for neutrally stable modes
of type II in a square duct $(A=1)$ subject to a vertical magnetic
field at various Hartmann numbers. }
\end{figure}

\begin{figure}
\begin{centering}
\includegraphics[width=0.5\textwidth]{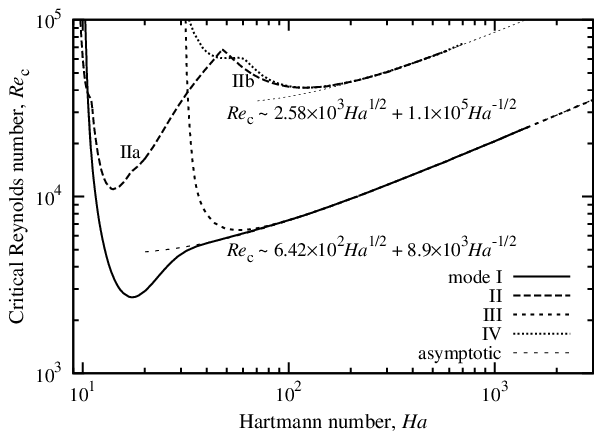}\put(-180,140){(\textit{a})}\includegraphics[width=0.5\textwidth]{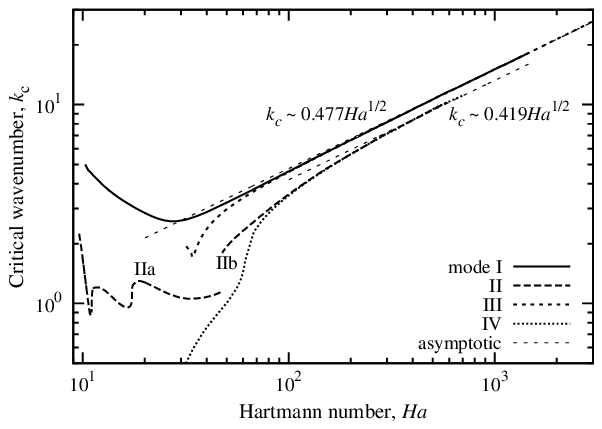}\put(-180,140){(\textit{b})}\\
\includegraphics[width=0.5\textwidth]{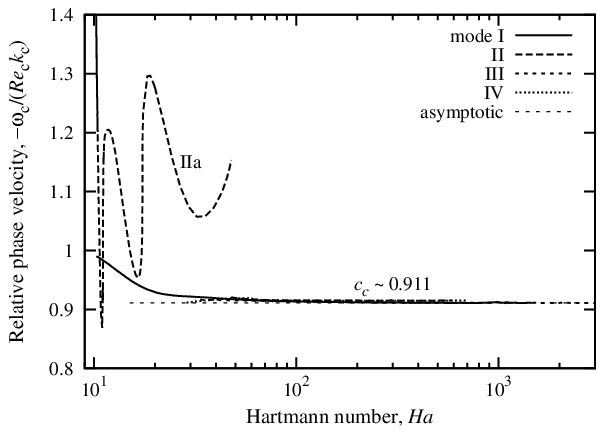}\put(-180,140){(\textit{c})} 
\par\end{centering}

\caption{\label{fig:rec_Ha_Ge1c}The critical Reynolds number (\emph{a}), wavenumber
(\emph{b}) and relative phase velocity (\emph{c}) against the Hartmann
number.}
\end{figure}

We start with a square duct, in which the flow without the magnetic
field is linearly stable \citep{TatYos90}. The magnetic field renders
the base flow linearly unstable at $\Ha\gtrsim9.6$ with respect to
a perturbation of symmetry type II. This perturbation is characterised
by the vorticity component along the magnetic field being an odd function
in the field direction and an even function spanwise. As shown in
the following, the anti-symmetric distribution along the field results
in a strong damping when the field strength is increased. The marginal
Reynolds number at which the maximum growth rate for this mode turns
zero $(\Re[\lambda]=0)$ is plotted in figure \ref{fig:rewk_Ge1_12}(a)
against the wavenumber for various Hartmann numbers. Besides the marginal
Reynolds number, neutrally stable perturbations are characterised
by their oscillation frequency $\omega=\Im[\lambda]$ and the associated
phase velocity $-\omega/k.$ It is useful to consider the latter relative
to the characteristic base flow velocity given by $\RE.$ This quantity
defined as $-\omega/(\RE k$) is subsequently referred to as the relative
phase velocity and shown in figure \ref{fig:rewk_Ge1_12}(b) for mode
II. Instability appears above the critical Reynolds number $\RE_{c}$,
which is defined by the global minimum on the neutral stability curve
for the respective Hartmann number. With the increase of $\Ha,$ the
critical Reynolds number for mode II in figure \ref{fig:rewk_Ge1_12}(a)
first quickly drops to a minimum $\RE_{c}\approx1.1\times10^{4}$
at $\Ha\approx14$ and then starts to increase. In some ranges of
the Hartmann number another minimum appears on the neutral stability
curve, which may cause the critical mode to jump from the first to
the second minimum as the latter becomes the global one. This switchover
between global minima shows up as a jump in both the critical wavenumber
and frequency, and as a break point in the dependence of the critical
Reynolds number on the Hartmann number. Such a jump is noticeable
in figure \ref{fig:rec_Ha_Ge1c} at $\Ha\approx48,$ where the instability
switches from mode IIa to IIb.

\begin{figure}
\begin{centering}
\includegraphics[width=0.5\textwidth]{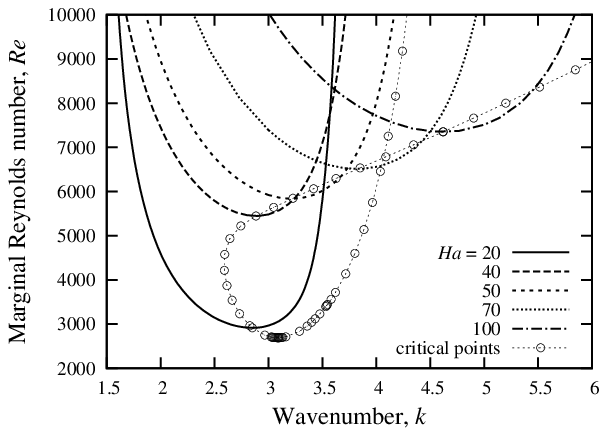}\put(-180,140){(\textit{a})}\includegraphics[width=0.5\textwidth]{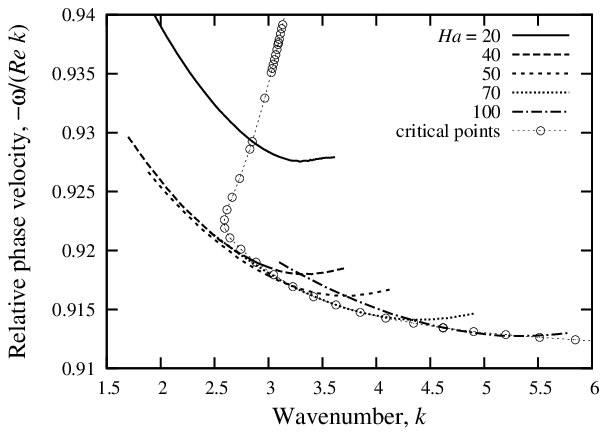}\put(-180,140){(\textit{b})}\\
\includegraphics[width=0.5\textwidth]{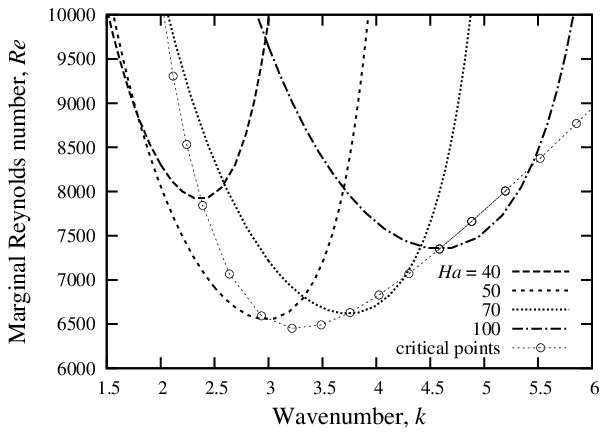}\put(-180,140){(\textit{c})}\includegraphics[width=0.5\textwidth]{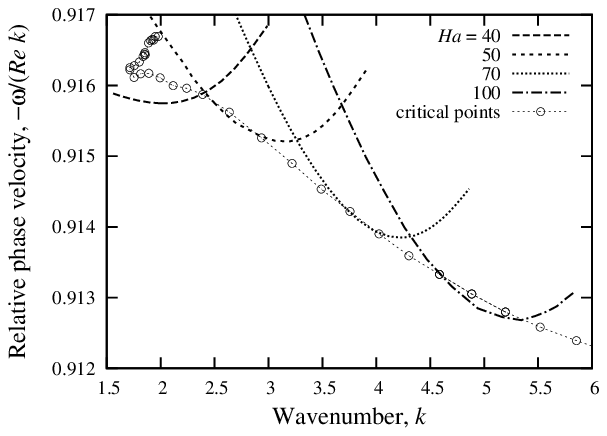}\put(-180,140){(\textit{d})} 
\par\end{centering}

\caption{\label{fig:rewk_Ge1_11-21}The marginal Reynolds number (\emph{a,c})
and the relative phase velocity (\emph{b,d}) versus the wavenumber
for neutrally stable modes of type I (\emph{a,b}) and type III (\emph{c,d}). }
\end{figure}

However, this jump is of secondary importance because a mode of type
I is seen in figure \ref{fig:rec_Ha_Ge1c}(a) to become more unstable
than mode II at $\Ha\approx10.6.$ Although mode I turns linearly
unstable at a slightly higher Hartmann number than mode II, its critical
Reynolds number decreases faster with the increase in the Hartmann
number than that for mode II. As seen from the neutral stability curves
in figure \ref{fig:rewk_Ge1_11-21}(a,b), with the increase of the
Hartmann number, the critical Reynolds number for mode I first quickly
drops to to a minimum $\RE_{c}\approx2697$ at $\Ha\approx17$ and
then starts to raise. With the increase of $\Ha$ mode I is quickly
approached from above by a mode of type III, which is seen in figure
\ref{fig:rec_Ha_Ge1c}(a) to appear at $\Ha\approx30$ and become
practically indistinguishable from the former at $\Ha\gtrsim80.$
There is also a mode of type IV appearing at $\Ha\approx30,$ which
in turn approaches mode II in a similar way as mode III approaches
mode I. Modes III/IV differ from modes I/II by the opposite symmetry
across the magnetic field. Namely, for modes III/IV, the vorticity
component along the magnetic field is an odd function in the spanwise
direction across the magnetic field, whereas it is an even function
for modes I/II. 

The most important feature of the instability seen in figure \ref{fig:rec_Ha_Ge1c}(\emph{a,b})
is the critical Reynolds number and the wavenumber for each of two
merged pairs of modes increasing in strong magnetic field as $\sim\Ha^{1/2}.$
The relative phase velocity shown in figure \ref{fig:rec_Ha_Ge1c}
tends to a constant $\sim0.911$ for both pairs of modes. This kind
of variation implies that in a strong magnetic field the instability
is determined by the internal length scale $\delta\sim\Ha^{-1/2},$
which is the characteristic thickness of the jets developing along
the walls parallel to the magnetic field. The base flow velocity correction
of order $\sim\Ha^{-1},$ which was discussed above, implies an $O(\Ha^{-1/2})$
correction to the critical Reynolds number. The best fit for modes
I/III yields \begin{eqnarray}
\RE_{c}(\Ha;A=1) & \approx & 642\Ha^{1/2}+8.9\times10^{3}\Ha^{-1/2},\label{eq:rec-I}\\
k_{c}(\Ha;A=1) & \approx & 0.477\Ha^{1/2},\label{eq:kc-I}\end{eqnarray}
 which are seen in figure \ref{fig:rec_Ha_Ge1c}(\emph{a,b}) to well
approximate numerical results for mode I down to $\Ha\approx30.$
Similarly, for modes II/IV, we find \begin{eqnarray}
\RE_{c}(\Ha;A=1) & \approx & 2580\Ha^{1/2}+1.1\times10^{5}\Ha^{-1/2},\label{eq:rec-II}\\
k_{c}(\Ha;A=1) & \approx & 0.419\Ha^{1/2},\label{eq:kc-II}\end{eqnarray}
 where the former is by nearly of a factor of four greater than (\ref{eq:rec-I}). 

\begin{figure}
\begin{centering}
\includegraphics[bb=100bp 95bp 290bp 255bp,clip,width=0.33\textwidth]{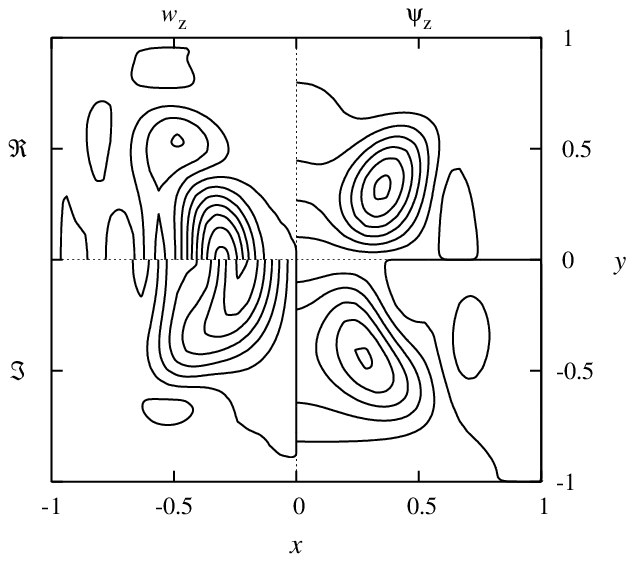}\put(-130,110){(\textit{a})}\includegraphics[bb=100bp 95bp 290bp 255bp,clip,width=0.33\textwidth]{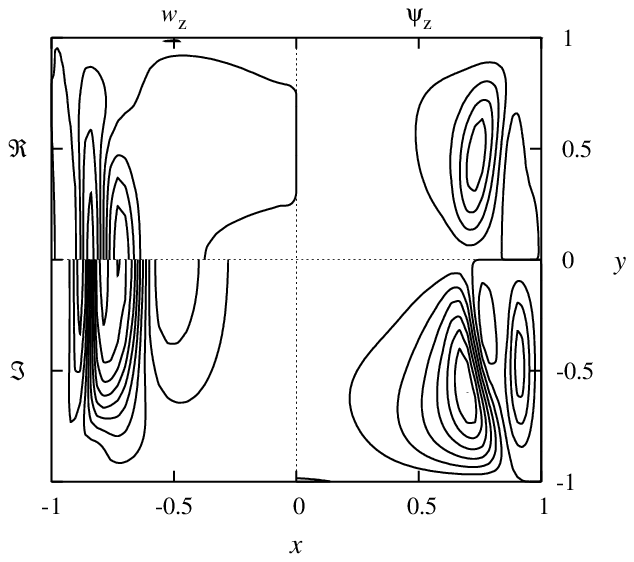}\put(-130,110){(\textit{b})}\includegraphics[bb=100bp 95bp 290bp 255bp,clip,width=0.33\textwidth]{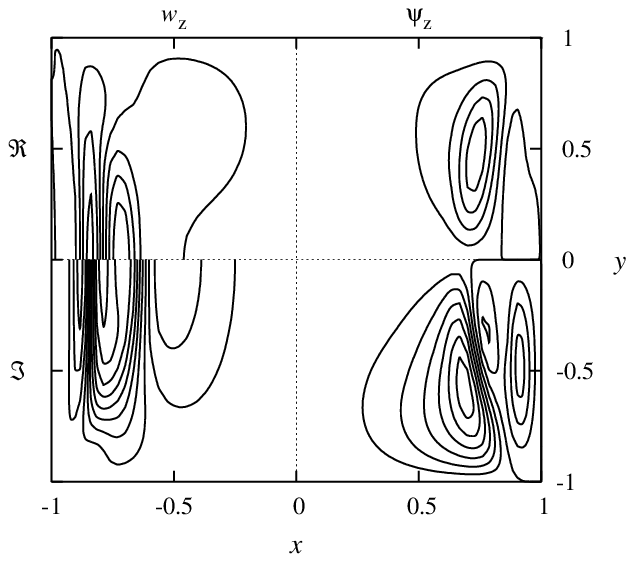}\put(-130,110){(\textit{c})}\\
\includegraphics[bb=100bp 95bp 290bp 255bp,clip,width=0.33\textwidth]{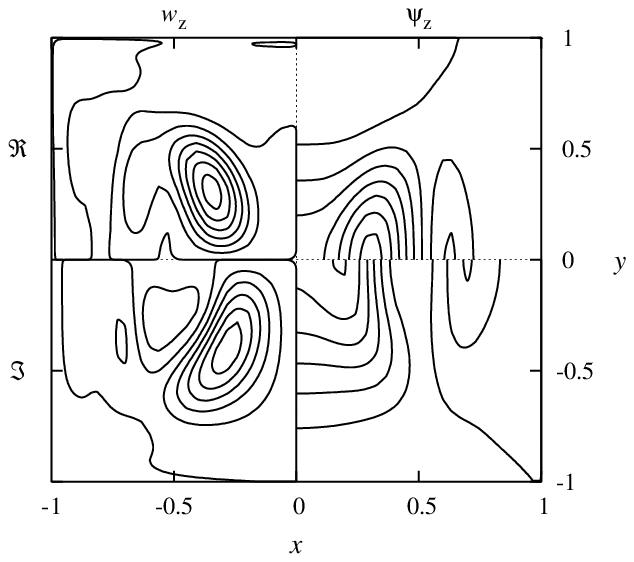}\put(-130,110){(\textit{d})}\includegraphics[bb=100bp 95bp 290bp 255bp,clip,width=0.33\textwidth]{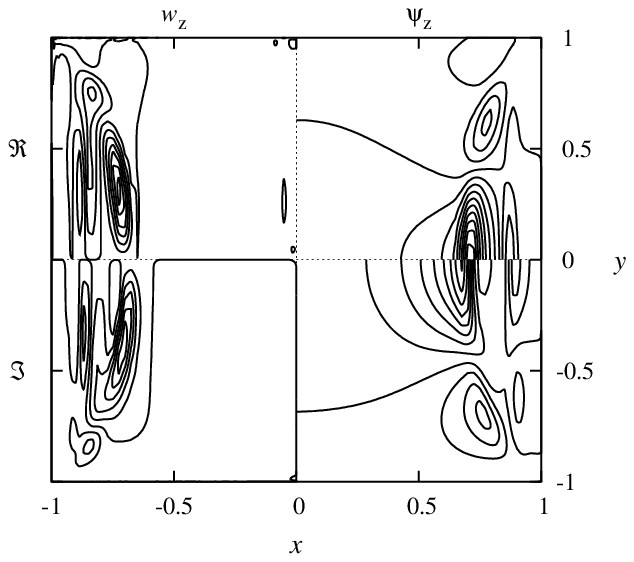}\put(-130,110){(\textit{e})}\includegraphics[bb=100bp 95bp 290bp 255bp,clip,width=0.33\textwidth]{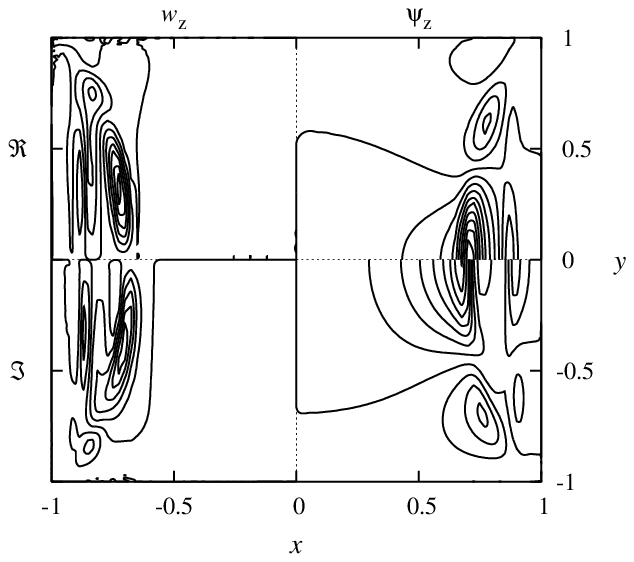}\put(-130,110){(\textit{f})}
\par\end{centering}

\caption{\label{fig:egv_Ge1_Ha15-100} Amplitude distributions of the real
$(y>0)$ and imaginary $(y<0)$ parts of $\hat{w}$ $(x<0)$ and $\hat{\psi}_{z}$
$(x>0)$ of the critical perturbations over one quadrant of duct cross-section
for instability modes I (\emph{a,b}), IIa (\emph{d,e}), III(\emph{c})
and IV(\emph{f}) at $\Ha=15$ (\emph{a,d}) and $\Ha=100$ (\emph{b,c,e,f}). }
\end{figure}

The instability threshold being nearly the same for the modes of opposite
spanwise symmetry in a sufficiently strong magnetic field implies
that the perturbations developing in the jets at the opposite walls
are effectively separated by the core region of the flow and, thus,
do not affect each other. This is confirmed by the patterns of the
critical perturbations, which are plotted over the duct cross-section
in figure \ref{fig:egv_Ge1_Ha15-100} for a moderate $(\Ha=15)$ and
a relatively strong $(\Ha=100)$ magnetic field. As shown in our previous
paper \citep{PAM10}, the flow perturbation can be represented by
the complex amplitudes of the streamwise $(z)$ component of velocity
$(\hat{w})$ and that of stream function $(\hat{\psi}_{z}),$ whose
isolines are plotted in the left $(x<0)$ and the right $(x>0)$ sides
of the cross-section, respectively. The real and imaginary parts of
perturbations plotted at the top and bottom halves of the cross-section
show the instant patterns shifted in time or in the stream-wise direction
by a quarter of period or wavelength, respectively. Although the real
and imaginary parts of the complex amplitude distributions completely
determine the evolution of perturbation over the harmonic oscillation
cycle, these two quantities are not uniquely defined. The main ambiguity
is due to the free choice of the initial time instant and the initial
stream-wise coordinate. This ambiguity can be partly eliminated by
choosing the phase of the complex velocity perturbation amplitude
so that \[
\int_{S}\Re[\vec{\hat{v}]}^{2}\, ds=\int_{S}\Im[\vec{\hat{v}]}^{2}\, ds,\]
where the integrals are taken over the duct cross-section $S.$ This
condition definines the phase up $\pm\pi/2$, which means swaping
the real and imaginary parts. The perturbation amplitude remains defined
up to a constant factor which is not important as the linear stability
theory predicts only the pattern but not the amplitude of the critical
perturbations.

Positive and negative values of $\hat{w}$ are respectively associated
with converging and diverging potential flow component in the cross-section
plane. The isolines of $\hat{\psi}_{z}$ correspond to the streamlines
of the solenoidal flow component in that plane. The critical perturbations
for modes I and II, which are shown in figures \ref{fig:egv_Ge1_Ha15-100}(\emph{a,b})
and (\emph{d,e}), respectively, differ by their vertical symmetry.
Namely, the perturbation of $\hat{w}$ and $\hat{\psi}_{z}$ are respectively
even and odd functions of $y$ for mode I, whereas they are odd and
even functions for mode II. Thus, the vortices for mode I rotate in
opposite senses in the upper and lower parts of the cross-section,
whereas for mode II there is one symmetric vortex spanning the whole
height of the duct. For both of these modes, the pairs of vortices
across the vertical mid-plane rotate in the same sense and, thus,
represent two parts of a bigger vortex spanning over the whole width
of the duct. At $\Ha=15,$ slightly above the Hartmann number at which
the flow turns linearly unstable, the critical perturbations are seen
in figure \ref{fig:egv_Ge1_Ha15-100}(\emph{a,d}) to be localised
close to the duct centre, where the two velocity maxima discussed
above first appear. In this case, the co-rotating vortices on the
opposite sides of the duct, whose symmetric half is shown at $x>0,$
are clearly connected by the flow through the vertical mid-plane.
However, this is no longer the case for a sufficiently strong magnetic
field. As seen in figure in \ref{fig:egv_Ge1_Ha15-100}(\emph{b,e}),
at $\Ha=100$ the critical perturbations for modes I and II are localised
at the side walls. Moreover, these perturbations have nearly the same
pattern as those for modes III and IV, which are shown in figure \ref{fig:egv_Ge1_Ha15-100}(\emph{c,f})
for the same $\Ha.$ It means that in a strong enough magnetic field
the perturbations at the opposite walls are effectively separated
by a stagnant core flow and, thus, do not affect each other. This
explains the merging of the instability thresholds for the modes with
the opposite spanwise symmetries seen in figure \ref{fig:rec_Ha_Ge1c}.

In weaker magnetic fields or in narrower ducts across the magnetic
field, which will be considered later, perturbations at the opposite
walls can either enhance or suppress each other depending on their
spanwise symmetry. The first is the case for the perturbations of
type I/II, which, as discussed above, have co-rotating vortices at
the opposite walls connected by a flow across the vertical mid-plane.
These perturbations are more unstable than those of type II/IV with
counter-rotating vortices at the opposite walls, which tend to suppress
each other, especially in moderate magnetic fields or in sufficiently
narrow ducts.

Besides the spatial amplitude distributions perturbations can be characterised
quantitatively by their kinetic energy distribution, which can be
represented in two different forms using either the velocity or vorticity/stream
function components \[
E\propto\int_{S}\hat{\left|\vec{v}\right|}^{2}\, ds=\int_{S}\Re[\hat{\vec{\omega}}\cdot\hat{\vec{\psi}}^{*}]\, ds,\]
 where $E$ is the kinetic energy of perturbation averaged over the
wavelength and the asterisk denotes the complex conjugate \citep{PAM10}.
At the moderate Hartmann number $\Ha=15$ considered above, most of
the kinetic energy, i.e. $60\%$ and $66\%$ for mode I and $88\%$
and $53\%$ for mode II, is carried by the $z$-component of velocity
and by the $y$-component of vorticity, respectively. For mode I,
next most energetic is the $x$-component of both velocity and vorticity,
which contain respectively about $27\%$ and $23\%$ of the energy.
For mode II, in this position are the $y$ velocity and the $x$ vorticity
components, which contain respectively about $10\%$ and $37\%$ of
the energy. 

\begin{figure}
\centering{}\includegraphics[bb=30bp 30bp 260bp 230bp,clip,width=0.5\textwidth]{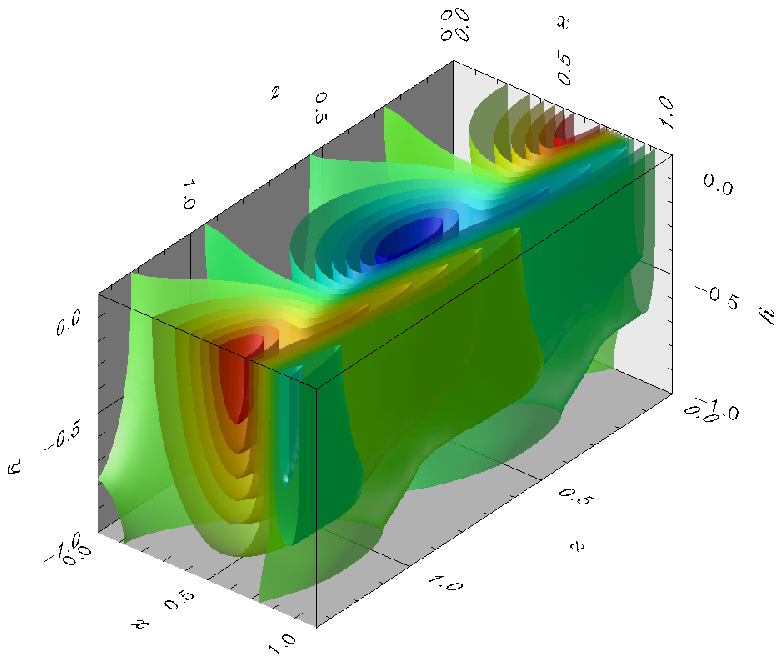}\put(-180,140){(\textit{a})}\includegraphics[bb=30bp 30bp 260bp 230bp,clip,width=0.5\textwidth]{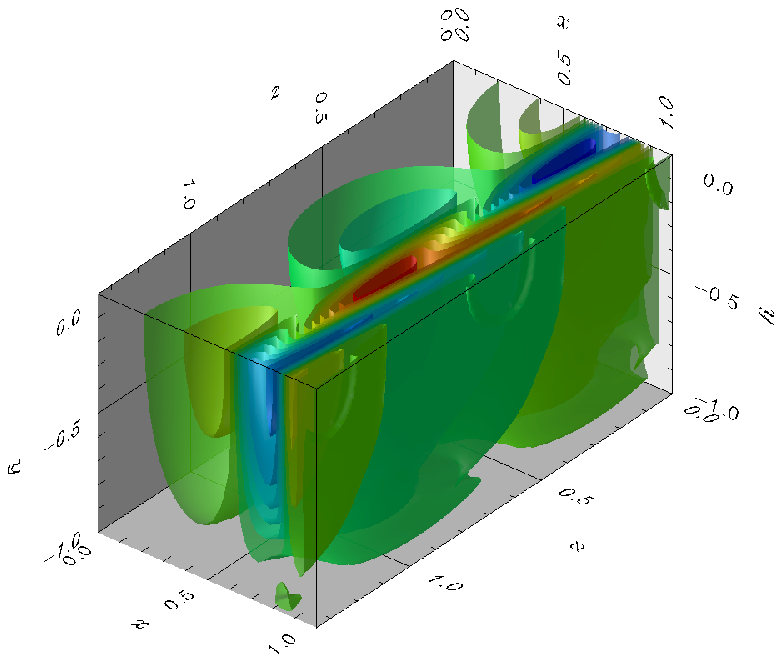}\put(-180,140){(\textit{b})}\caption{\label{fig:wz_Ge1_Ha100} Isosurfaces of $\hat{\psi}_{y}$ (a) and
longitudinal velocity $\hat{w}$ (b) perturbations over wavelength
in one quadrant of the duct cross-section for instability mode I at
$\Ha=100$ in a vertical magnetic field. }
\end{figure}

The energy distribution for the most unstable modes I and II becomes
simpler in a strong magnetic field. For example, at $\Ha=100,$ $86\%$
of the energy for these modes is concentrated in the $y$-component
of vorticity, while the rest is distributed nearly equally between
the two other vorticity components. This component of vorticity is
associated with the circulation in the $(x,z)$-planes transverse
to the magnetic field. The streamlines of this circulation are represented
by the isolines of $\psi_{y},$ whose spatial distribution is shown
in figure \ref{fig:wz_Ge1_Ha100}(\emph{a}). Note that the distribution
of $\psi_{y}$ is very close to that of the electric potential $\phi$
because the equations (\ref{eq:phih}) and (\ref{eq:psih}) governing
both quantities are the same. Moreover, $\psi_{y}$ and $\phi$ satisfy
the same boundary condition at the wall parallel to the magnetic field.
Thus, both quantities differ only in the vicinity of the wall normal
to the magnetic field, where they have different boundary conditions.

The transverse character of circulation for modes I and III is confirmed
also by the kinetic energy distribution over the velocity components.
Only about $9\%$ of the energy is carried by the velocity component
along the magnetic field, while $67\%$ by the streamwise $(z)$ velocity
perturbation, whose spatial distribution is shown in \ref{fig:wz_Ge1_Ha100}(b).
The relatively low contribution of the spanwise $(x)$ velocity component,
which carries the remaining $24\%$ of the energy, is due to the relatively
long wavelength of perturbation $\lambda_{c}=2\pi/k_{c},$ which according
to (\ref{eq:kc-I}) is by approximately a factor of 7 larger than
the thickness of the jet (\ref{eq:delta}).

In a strong magnetic field, the energy distribution in modes II/IV
is essentially different from that in modes I/II considered above.
Namely, the latter two have only $5-6\%$ of their energy in the spanwise
$(x)$ velocity component, which implies a circulation constrained
mainly to the $(y,z)$-planes parallel to the side walls. Similar
to the previous two modes, $66\%$ of the energy is carried by the
streamwise velocity component. Although circulation mostly occurs
in the $(y,z)$-planes, only $19\%$ of the energy is contained the
transverse $(x$) vorticity/stream function component, while $56\%$
are still contained in the vorticity/stream function component along
the magnetic field. This scatter of energy between the vorticity components
is due to the confinement of circulation in narrow layers parallel
to the side walls. The confinement causes a strong variation of the
velocity perturbation in the spanwise $(x)$ direction and, thus,
produces a significant vorticity components tangential to the plane
of circulation. 

Finally, we consider the effect of the duct aspect ratio on the instability
threshold of the two most unstable modes. In order to generalise the
above results for square duct to arbitrary aspect ratios it is instructive
to start with a horizontal magnetic field, which is directed along
the fixed dimension of the duct, i.e. the width in our case. In sufficiently
strong horizontal magnetic field, the variation of the instability
threshold with the aspect ratio turns out to be particularly simple
and directly related to that of the square duct considered above.
We shall use this fact and the similarity of horizontal and vertical
magnetic field orientations to obtain more general relations for the
latter orientation.

It is important to notice that the thickness of parallel layers $\delta$
defined by (\ref{eq:delta}), which is the characteristic length scale
of the instability, varies not only with the Hartmann number but also
with the aspect ratio $A$, which defines the size of the duct along
the magnetic field. When the magnetic field is directed horizontally
along the fixed dimension of the duct, $\delta$ becomes independent
of $A$ and varies only with $\Ha$ as in the case of square duct.
This simplification is our main motivation for considering first horizontal
magnetic field.  

\begin{figure}
\begin{centering}
\includegraphics[width=0.5\textwidth]{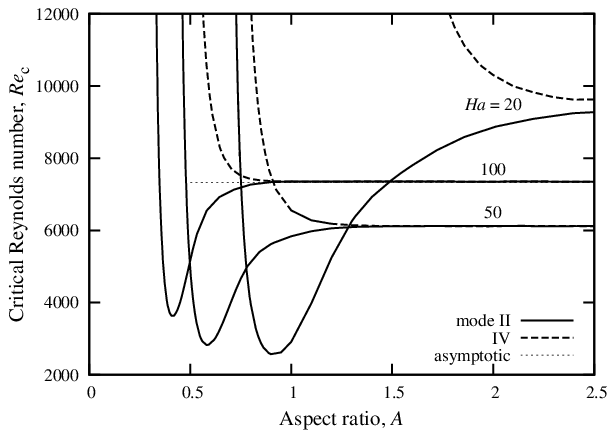}\put(-180,140){(\textit{a})}\includegraphics[width=0.5\textwidth]{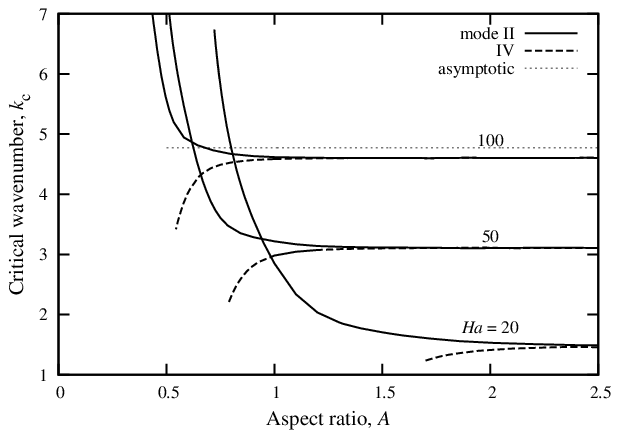}\put(-180,140){(\textit{b})}\\
\includegraphics[width=0.5\textwidth]{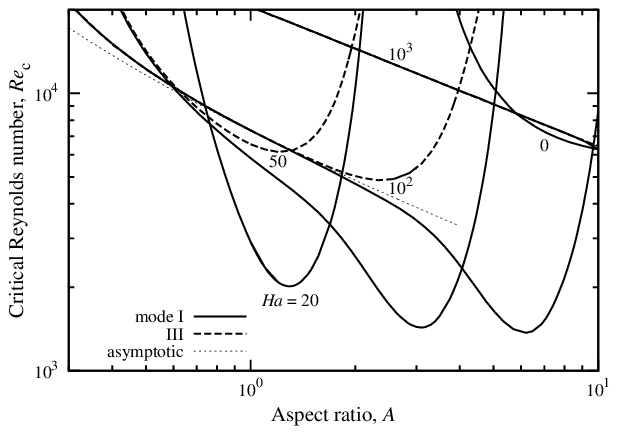}\put(-180,140){(\textit{c})}\includegraphics[width=0.5\textwidth]{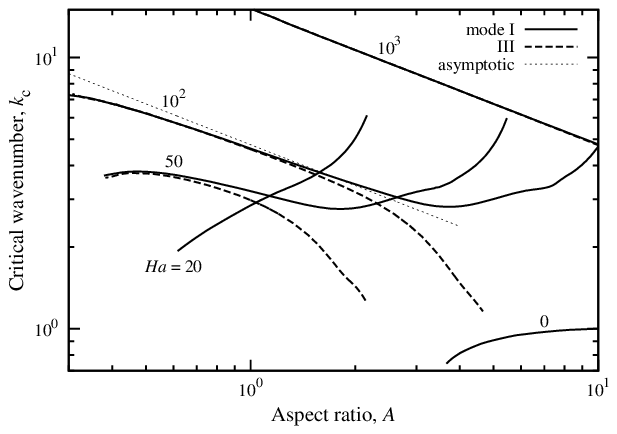}\put(-180,140){(\textit{d})} 
\par\end{centering}

\caption{\label{fig:Rec-A}The critical Reynolds number (a,c) and the wavenumber
(b,d) versus the aspect ratio for modes II/IV in a horizontal magnetic
field (a,b) and for modes I/III in a vertical magnetic field (c,d)
at various Hartmann numbers. }
\end{figure}

On changing the magnetic field from vertical to horizontal, modes
I/III swap with II/IV, which, thus, become the most unstable ones.
For sufficiently large aspect ratios, the critical Reynolds number
and the wavenumber are seen in figure \ref{fig:Rec-A}(\emph{a,b})
be the same for both modes, which agree with the strong field asymptotics
(\ref{eq:rec-I},\ref{eq:kc-I}). As discussed above, this implies
that the instabilities developing in the jets at the parallel walls
are effectively separated by a stagnant core of the flow and, thus,
do not affect each other. It changes at small aspect ratios, when
the walls parallel to the magnetic field are sufficiently close to
each other. Then the vortices at the opposite walls start to interact,
which causes the thresholds for both modes to diverge. For mode IV,
the vortices at the opposite walls counter-rotate and, thus, tend
to suppress each other, which results in the stabilisation of the
flow. It is the opposite for mode II, whose instability threshold
first drops as the co-rotating vortices at the opposite walls start
to enhance each other. However, with further reduction of the aspect
ratio the critical Reynolds number attains a minimum and then starts
to increase following that for mode IV. The raise of $\RE_{c}$ for
mode II is associated with the increase of the critical wavenumber.
This corresponds to the reduction of wavelength which is required
for the vortices to fit between closely spaced parallel walls.

Now we turn to vertical magnetic field, which is oriented along the
variable height of the duct. This slightly more complicated case can
be reduced to the previous one by taking the height of the duct as
the length scale and rotating the duct by 90 degrees. This is equivalent
to the substitutions $A'=1/A,$ $\RE'=A\RE,$ $\Ha'=A\Ha,$ and $k'=Ak,$
where the parameters with primes correspond to the case of horizontal
magnetic field considered above. Then we can use the result for horizontal
magnetic field found above according to which the critical parameters
in strong magnetic field approach those for a square duct given by
(\ref{eq:rec-I},\ref{eq:kc-I}). Substituting the primed parameters
into (\ref{eq:rec-I},\ref{eq:kc-I}) we obtain\begin{eqnarray}
\RE_{c}(\Ha;A) & = & A^{-1}\RE_{c}(A\Ha;1),\label{eq:Rec-A}\\
k_{c}(\Ha;A) & = & A^{-1}k_{c}(A\Ha;1),\label{eq:kc-A}\end{eqnarray}
which are seen in figure \ref{fig:Rec-A}(c,d) to fit the numerical
results well in the intermediate range of aspect ratios, where the
thresholds for both most unstable modes I/III merge. In this range
both the critical Reynolds number and wavenumber for a fixed $\Ha$
vary asymptotically with the aspect ratio as $\sim A^{-1/2}.$ This
variation, which is due to the increase of the jet width (\ref{eq:delta})
as $\sim A^{1/2},$ breaks down at both small and large $A.$ In the
former limit, asymptotic relations (\ref{eq:Rec-A},\ref{eq:kc-A})
turn inapplicable because the effective Hartmann number $A\Ha$ based
on the height becomes too small for (\ref{eq:rec-I},\ref{eq:kc-I})
to be valid. In the latter limit, the jets become so wide that the
vortices at the opposite walls start to interact resulting in the
divergence of the instability thresholds and the eventual stabilisation
described above for the case of horizontal magnetic field. 

\section{\label{sec:sum}Discussion and conclusions}

We have presented numerical results concerning linear stability of
a liquid metal flow in a rectangular duct with perfectly electrically
conducting walls subject to a uniform transverse magnetic field. It
was found that a linearly stable flow in a square duct turns unstable
as a relatively weak magnetic field with the Hartmann number $\Ha\approx9.6$
is applied. The instability is due to two weak jets, which first appear
at the centre of the duct and then move to the walls parallel to the
magnetic field as the field strength is increased. The instability
follows the jets and in a sufficiently strong magnetic field becomes
confined in the layers of characteristic thickness $\delta\sim\Ha^{-1/2}$
located at the parallel walls. The thickness $\delta$ determines
the characteristic length scale of the instability, which results
in both the critical Reynolds and wave numbers scaling as $\sim\delta^{-1}.$
Owing to the double reflection symmetry of the problem, perturbations
with four different symmetry combinations along and across the magnetic
field decouple from each other and, thus, are considered separately.
The most unstable is a pair of perturbations with an even distribution
of the vorticity along the magnetic field. These two modes represent
strongly non-uniform vortices aligned with the magnetic field, which
rotate either in the same or opposite directions across the magnetic
field. The former enhance while the latter weaken one another provided
that the magnetic field is not too strong or the walls parallel to
the field are not too far apart. In a strong magnetic field, when
the vortices at the opposite walls are well separated from the core
flow, the critical Reynolds number and wavenumbers for both of these
instability modes are the same: $\RE_{c}\approx642\Ha^{1/2}+O(\Ha^{-1/2})$
and $k_{c}\approx0.477\Ha^{1/2}.$ The other pair of critical perturbations,
which differs from the previous one by an odd distribution of vorticity
along the magnetic field, is more stable with approximately four times
higher critical Reynolds number.

The basic instability characteristics described above resemble those
for the Hunt's flow, which has a similarly increasing, however a significantly
lower, critical Reynolds number $\RE_{c}\sim91\Ha^{1/2}$ \citep{PAM10}.
The difference becomes substantial when the average rather than the
maximum velocity is considered. Namely, the critical Reynolds number
based on the average velocity for this flow increases in a strong
magnetic field in the same way as $\bar{\RE}_{c}\approx519\Ha^{1/2},$
whereas it tends to a constant $\bar{\RE}_{c}\approx112$ for Hunt's
flow. This asymptotically constant $\RE_{c}$ is due to a principally
different velocity distribution in Hunt's flow with the jets of thickness
$\sim\Ha^{-1/2}$ dominating the flow rate. Constant $\bar{\RE}_{c}\approx313$
has been found by \citet{Tin-etal91} for the linear stability of
the flow in a square duct with thin but relatively well-conducting
walls in a strong transverse magnetic field. This flow represents
an intermediate case in terms of the conductivity of parallel wall
between the perfectly conducting one considered in this study and
the insulating one for Hunt's flow. Although the leading-order velocity
distribution considered by \citet{Tin-etal91} is nearly the same
as that of Hunt's jet, there is one principal difference between two.
Namely, in a duct with thin parallel walls of a moderate conductance
ratio $c=\sigma_{w}d_{w}/\sigma d,$ where $\sigma_{w}$ and $d_{w}$
are the electrical conductivity and the thickness of wall, satisfying
$\Ha^{-1/2}\ll c\ll\Ha^{1/2},$ jets carry a volume flux of the same
order of magnitude as that of the core flow, whereas the contribution
of the latter is negligible in Hunt's flow. In a square duct with
thin walls, in which the core flow in strong magnetic field carries
$3/4$ of the total volume flux \citep{Wal81}, the jet velocity is
by a factor of about four lower than that for Hunt's flow at the same
$\bar{\RE.}$ It means that jets are more unstable at wall of finite
conductivity than in both limiting cases of insulating and perfectly
conducting parallel walls. 

The experiment best matching the model considered in this study has
been carried out by \citet{BraGel68}, who measured a flow of mercury
 at $\bar{\RE}\approx4\times10^{4}$ and $\Ha=174$ in a rectangular
copper duct with a relatively high wall conductance ratio, which was
$20$ and $10$ for the Hartmann and the parallel walls, respectively.
The magnetic field was applied transversely to the longest side of
the duct with the aspect ratio of $1.5.$ First, the authors found
a maximum velocity in jets exceeding the theoretical prediction for
perfectly conducting duct by approximately $50\%.$ At this Hartmann
number, we find that $19\%$ of the excess jet velocity may be due
to the finite wall conductivity, when the latter is included in the
numerical solution. Note that the effect of imperfectly conducting
walls is not entirely determined by the relative conductance of the
parallel layers $c\Ha^{1/2}$ as it may appear from \citet{Hunt65}
\citep[see also][p. 146]{MulBuh01}. Namely, $c\Ha^{1/2}\gg1$ means
only that the parallel walls cannot be treated as insulating $(c\not=0)$.
But it does not necessary mean that the walls may be assumed perfectly
conducting $(c=\infty).$ As shown by \citet{Wal81}, the latter approximation
requires a much higher wall conductance ratio $c\Ha^{-1/2}\gg1.$
Thus, the effect of imperfectly conducting walls increases rather
than decreases with the field strength, which results in the jet velocity
relative to that of the core increasing with the field strength as
$\sim\Ha^{1/2}/c$ \citep{Wal81}. However, the most significant deviation
from the laminar flow solution by \citet{Hunt65} was the jet thickness,
which was found by \citet{BraGel68} to be several times greater than
expected. Such a broadening of jets is likely to be due to the turbulence
which is expected in the experiment at the Reynolds number significantly
above the linear stability threshold $\bar{\RE}_{c}\approx6850$ predicted
by our analysis for this setup.

In conclusion, note that when the distance of the velocity maximum
from the parallel wall (\ref{eq:delta}) is taken as the length scale,
(\ref{eq:rec-I}) becomes $\RE_{c}^{\delta}\approx1230.$ This small
critical Reynolds number may be due to the inviscid nature of the
instability caused by the inflection point of the base velocity profile.
It also implies that the instability may be supercritical rather than
subcritical as for typical shear flows. The supercritical character
of instability may explain why it appears significantly above the
linear stability threshold as recently reported by \citet{KinKnaMol09}.
They observed small-amplitude vortices in the jets at the parallel
walls for $2500\leq\bar{\RE}\leq3700$ at $\Ha=200$ in the numerical
simulation of a flow in a rectangular duct with thin walls. These
vortices were found to be subcritically unstable for $3500\leq\bar{\RE}\leq3700,$
which may correspond to the large amplitude instabilities observed
in the experiments \citep{ReePic89,BurBarMulTsi00}. If the instability
in perfectly conducting duct like that in ducts with thin walls of
finite conductivity is supercritical, it is expected to appear above
the absolute instability threshold which cannot be lower than that
of the convective instability predicted by the classical linear stability
analysis used in this study \citep{PG97}. The absolute instability
threshold for these jet-type flows, at which small-amplitude self-sustained
vortices are expected to appear, is not yet known. 

\begin{acknowledgements}

S.A. is grateful to Leverhulme Trust for financial support of this
work. The authors are indebted to the Faculty of Engineering and Computing
of Coventry University for the opportunity to use its high performance
computer cluster.

\end{acknowledgements}
\appendix

\section{\label{sec:app}Vector stream function/vorticity formulation}

We use the vector stream function $\vec{\psi},$ which is introduced
to satisfy the incompressiblity constraint $\vec{\nabla}\cdot\vec{v}=0$
for the flow perturbation by seeking the velocity distribution in
the form $\vec{v}=\vec{\nabla}\times\vec{\psi}.$ Since the velocity
is invariant upon adding the gradient of arbitrary function to $\vec{\psi},$
we can impose an additional constraint \begin{equation}
\vec{\nabla}\cdot\vec{\psi}=0,\label{eq:divpsi}\end{equation}
which is analogous to the Coulomb gauge for the magnetic vector potential
$\vec{A}$ \citep{Jack98}. Similarly to the incompressiblity constraint
for $\vec{v},$ this gauge leaves only two independent components
of $\vec{\psi}.$ 

The pressure gradient is eliminated by applying \textit{curl} to (\ref{eq:NS}).
This yields two dimensionless equations for $\vec{\psi}$ and $\vec{\omega}$
\begin{eqnarray}
\partial_{t}\vec{\omega} & = & \vec{\nabla}^{2}\vec{\omega}-\RE\vec{g}+\Ha^{2}\vec{h},\label{eq:omeg}\\
0 & = & \vec{\nabla}^{2}\vec{\psi}+\vec{\omega},\label{eq:psi}\end{eqnarray}
 where $\vec{g}=\vec{\nabla}\times(\vec{v}\cdot\vec{\nabla})\vec{v},$
and $\vec{h}=\vec{\nabla}\times\vec{f}$ are the \textit{curls} of
the dimensionless convective inertial and electromagnetic forces,
respectively. 

The boundary conditions for $\vec{\psi}$ and $\vec{\omega}$ are
obtained as follows. The impermeability condition applied integrally
as $\int_{s}\vec{v}\cdot\vec{ds}=\oint_{l}\vec{\psi}\cdot\vec{dl}=0$
to an arbitrary area of wall $s$ encircled by a contour $l$ yields
$\left.\psi_{\tau}\right|_{s}=0.$ This boundary condition substituted
into (\ref{eq:divpsi}) results in $\left.\partial_{n}\psi_{n}\right|_{s}=0.$
In addition, the no-slip condition applied integrally $\oint_{l}\vec{v}\cdot\vec{dl}=\int_{s}\vec{\omega}\cdot\vec{ds}$
yields $\left.\omega_{n}\right|_{s}=0.$

Linear stability of the base flow $\{\bar{\vec{\psi}},\bar{\vec{\omega}},\bar{\phi}\}(x,y)$
is analysed with respect to infinitesimal disturbances in the form
of harmonic waves travelling along the axis of the duct \[
\{\vec{\psi},\vec{\omega},\phi\}(\vec{r},t)=\{\bar{\vec{\psi}},\bar{\vec{\omega}},\bar{\phi}\}(x,y)+\{\hat{\vec{\psi}},\hat{\vec{\omega}},\hat{\phi}\}(x,y)e^{\gamma t+\i kz},\]
 where $k$ is a wavenumber and $\gamma$ is, in general, a complex
growth rate. This expression substituted into (\ref{eq:omeg},\ref{eq:psi})
results in \begin{eqnarray}
\gamma\hat{\vec{\omega}} & = & \vec{\nabla}_{k}^{2}\hat{\vec{\omega}}-\RE\hat{\vec{g}}+\Ha^{2}\hat{\vec{h}},\label{eq:omegh}\\
0 & = & \vec{\nabla}_{k}^{2}\hat{\vec{\psi}}+\hat{\vec{\omega}},\label{eq:psih}\\
0 & = & \vec{\nabla}_{k}^{2}\hat{\phi}-\hat{\omega}_{\shortparallel},\label{eq:phih}\end{eqnarray}
 where $\vec{\nabla}_{k}\equiv\vec{\nabla}_{\perp}+\i k\vec{e}_{z};$
$\shortparallel$ and $\perp$ respectively denote the components
along and transverse to the magnetic field in the $(x,y)$-plane.
Because of the solenoidality of $\hat{\vec{\omega}},$ we need only
the $x$- and $y$-components of (\ref{eq:omegh}), which contain
$\hat{h}_{\perp}=-\partial_{xy}\hat{\phi}-\partial_{\shortparallel}\hat{w},$
$\hat{h}_{\shortparallel}=-\partial_{\shortparallel}^{2}\hat{\phi}$
and \begin{eqnarray}
\hat{g}_{x} & = & k^{2}\hat{v}\bar{w}+\partial_{yy}(\hat{v}\bar{w})+\partial_{xy}(\hat{u}\bar{w})+\i2k\partial_{y}(\hat{w}\bar{w}),\label{eq:gx}\\
\hat{g}_{y} & = & -k^{2}\hat{u}\bar{w}-\partial_{xx}(\hat{u}\bar{w})-\partial_{xy}(\hat{v}\bar{w})-\i2k\partial_{x}(\hat{w}\bar{w}),\label{eq:gy}\end{eqnarray}
 where \begin{eqnarray*}
\hat{u} & = & \i k^{-1}(\partial_{yy}\hat{\psi}_{y}-k^{2}\hat{\psi}_{y}+\partial_{xy}\hat{\psi}_{x}),\\
\hat{v} & = & -\i k^{-1}(\partial_{xx}\hat{\psi}_{x}-k^{2}\hat{\psi}_{x}+\partial_{xy}\hat{\psi}_{y}),\\
\hat{w} & = & \partial_{x}\hat{\psi}_{y}-\partial_{y}\hat{\psi}_{x}.\end{eqnarray*}
 The relevant boundary conditions are\begin{eqnarray}
\hat{\phi}=\hat{\psi}_{y}=\partial_{x}\hat{\psi}_{x}=\partial_{x}\hat{\psi}_{y}-\partial_{y}\hat{\psi}_{x}=\hat{\omega}_{x}=0 & \mbox{at} & x=\pm1,\label{eq:bndcx}\\
\hat{\phi}=\hat{\psi}_{x}=\partial_{y}\hat{\psi}_{y}=\partial_{x}\hat{\psi}_{y}-\partial_{y}\hat{\psi}_{x}=\hat{\omega}_{y}=0 & \mbox{at} & y=\pm A.\label{eq:bndcy}\end{eqnarray}

\end{document}